\documentclass[12pt]{article}

\usepackage[utf8]{inputenc}
\usepackage[T1]{fontenc}
\usepackage{geometry}
\usepackage{setspace}
\usepackage[round]{natbib}
\usepackage{hyperref}
\usepackage{booktabs}
\usepackage{longtable}
\usepackage{array}
\usepackage{graphicx}
\usepackage{titlesec}
\usepackage{xcolor}
\usepackage{tcolorbox}
\usepackage{parskip}
\usepackage{float}
\usepackage{enumitem}

\definecolor{abstractbg}{RGB}{237,241,247}
\definecolor{linkblue}{RGB}{0,102,204}

\hypersetup{
    colorlinks=true,
    linkcolor=linkblue,
    urlcolor=linkblue,
    citecolor=linkblue
}

\titleformat{\section}{\normalfont\Large\bfseries}{\thesection}{1em}{}
\titleformat{\subsection}{\normalfont\large\bfseries}{\thesubsection}{1em}{}
\titlespacing*{\section}{0pt}{1.5ex plus 0.5ex minus 0.2ex}{1ex plus 0.2ex}
\titlespacing*{\subsection}{0pt}{1.2ex plus 0.3ex minus 0.2ex}{0.8ex plus 0.2ex}

\setlist{nosep, leftmargin=*}

\title{}
\author{}
\date{}

\begin{document}

\begin{tcolorbox}[
    colback=abstractbg,
    colframe=abstractbg,
    arc=4pt,
    boxrule=0pt,
    left=20pt,
    right=20pt,
    top=20pt,
    bottom=15pt
]

\begin{center}
{\fontsize{18}{22}\selectfont\bfseries Attachment Styles and AI Chatbot Interactions\\Among College Students}
\end{center}

\vspace{1em}

\textbf{Ziqi Lin}\textsuperscript{1,*}, \textbf{Taiyu Hou}\textsuperscript{1,*}

\vspace{0.3em}
\textsuperscript{1}Department of Applied Psychology, New York University

\vspace{1em}

The use of large language model (LLM)-based AI chatbots among college students has increased rapidly, yet little is known about how individual psychological attributes shape students' interaction patterns with these technologies. This qualitative study explored how college students with different attachment styles describe their interactions with ChatGPT. Using semi-structured interviews with seven undergraduate students and grounded theory analysis, we identified three main themes: (1) AI as a low-risk emotional space, where participants across attachment styles valued the non-judgmental and low-stakes nature of AI interactions; (2) attachment-congruent patterns of AI engagement, where securely attached students integrated AI as a supplementary tool within their existing support systems, while avoidantly attached students used AI to buffer vulnerability and maintain interpersonal boundaries; and (3) the paradox of AI intimacy, capturing the tension between students' willingness to disclose personal information to AI while simultaneously recognizing its limitations as a relational partner. These findings suggest that attachment orientations play an important role in shaping how students experience and interpret their interactions with AI chatbots, extending attachment theory to the domain of human--AI interaction.

\vspace{1em}

\begin{minipage}[b]{0.70\textwidth}
\small
\textbf{Correspondence:} \href{mailto:zl4492@nyu.edu}{zl4492@nyu.edu}, \href{mailto:th2902@nyu.edu}{th2902@nyu.edu}\\[0.2em]
\textbf{Note:} *These authors contributed equally to this work.
\end{minipage}%
\hfill
\begin{minipage}[b]{0.27\textwidth}
\raggedleft
\includegraphics[height=2cm]{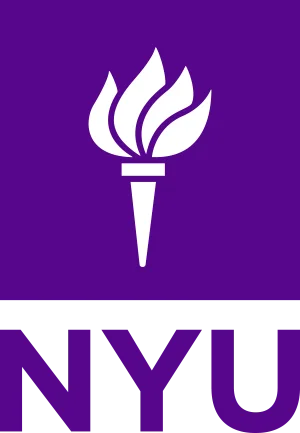}
\end{minipage}

\end{tcolorbox}

\vspace{1.2em}

\section{Introduction}

In recent years, the use of artificial intelligence (AI) tools, like ChatGPT, among college students has increased rapidly amidst technological advancement. A survey of more than 1,000 college undergraduate students found that the proportion of students who use ChatGPT to support their academic performance has increased from 53\% to 88\% since 2024---a significant increase representing the vast majority of students \citep{freeman2025}. ChatGPT is a large language model-based (LLM) AI chatbot, characterized by its ability to generate responses that mimic natural language processing in conversations \citep{yigci2024}. Previous research has found that students use LLM-based AI chatbots particularly for academic-related purposes, such as time management for studying, understanding complex topics, or improving overall academic performance \citep{morell2025}.

Although existing research has largely focused on understanding the functional use and benefits of AI among college students, emerging evidence also suggests that many students have relational interactions with AI in emotionally meaningful ways---such as improving motivation, emotional regulation, communication skills, and overall mental well-being \citep{lin2025}. Many students describe AI chatbots as comforting, nonjudgmental, or even ``friend-like,'' particularly during stressful moments \citep{brandtzaeg2022, skjuve2021, zhang2025}. Further, use of ChatGPT in academic settings is positively correlated with higher happiness scores due to reduced workload, stress, and more clearly articulated ideas \citep{klimova2025}. Overall, past research has categorized ChatGPT usage among students in three main ways: academic support, personalized learning experience, and skill development \citep{labadze2023}. These findings suggest that AI chatbots can serve both functional and emotional purposes, but the ways and extent to which college students engage with AI differ widely.

Despite growing interest in the use of AI in higher education settings, less is known about how individual psychological attributes shape students' interaction patterns with AI---the extent to which they disclose personal information, trust AI-generated responses, and perceive human likeness of ChatGPT. Prior studies often focus on examining AI use frequency and emotional or academic outcomes, but insights on students' personal experience and their interpretations of interaction patterns with AI remain limited. Although frequent and emotionally charged AI chatbot use has been found to be associated with higher stress and low confidence, these findings have not yet been connected to a broader theoretical framework that explains why such differences emerge \citep{saglam2025, zhang2025}.

Attachment theory may be used to explain the individual differences in AI usage. Research shows that attachment styles play a key role in individuals' social processing within human-human interactions, where securely attached individuals tend to have better theory of mind (ToM) skills and higher trust in social relationships compared to those with avoidant attachment styles \citep{bao2022, gallistl2024}. Recent work shows that attachment tendencies can also extend to non-human agents such as pets and digital systems \citep{ciacchella2024, zilcha2011}. Similarly, because chatbots may provide support or a nonjudgmental space for expression \citep{jiang2022, skjuve2021}, attachment processes may also arise in human--AI interactions. However, little research has examined how students with different attachment styles interact with AI chatbots differently. To address this gap, the present study seeks to explore how college students with different attachment styles describe their interactions with ChatGPT, an LLM-based AI chatbot.

\subsection{Individual Differences in AI Chatbot Interaction}

College students' reliance on LLM-based AI chatbots has been on the rise amid rapid technological advancement and increased adoption of AI for academic purposes \citep{fageeh2025, saglam2025}. College students display substantial variability in how they interact with AI chatbots, where a large proportion primarily rely on AI for functional purposes to support academic tasks, most commonly on explaining concepts, summarizing articles, and generating research ideas \citep{freeman2025}. Students often perceive AI chatbots as an unprecedented and advanced tool capable of producing structured explanations, code, statistical analysis, and even entire sections of scientific text \citep{macdonald2023}, thereby supporting efficiency, time-saving, and improved understanding of complex topics \citep{saglam2025}.

At the same time, emotional or companionship-driven engagement with AI chatbots is also widespread, in that students value the nonjudgmental or even friend-like experience when having a conversation with AI \citep{brandtzaeg2022, freeman2025}. Interestingly, students tend to underestimate their emotional engagement with AI---a recent study found that 92.9\% displayed companionship-style behaviors in actual conversation logs, while only 11.8\% explicitly acknowledged using AI for companionship \citep{zhang2025}. These findings highlight how relational and emotionally expressive interaction patterns can emerge even among students who consider themselves primarily task-oriented users \citep{brandtzaeg2022, zhang2025}. Moreover, qualitative studies show that emotional attachment and affective dependence on chatbots may develop gradually, sometimes accompanying tendencies such as emotional overreliance or reduced real-life social engagement \citep{saglam2025, xie2022}.

Some students display mixed-use patterns that merge academic efficiency and emotional regulation. Studies show that students describe using AI for explaining difficult concepts, in addition to using AI as support against stress or to regain a sense of control in challenging academic times \citep{freeman2025, saglam2025}. This dual role of AI technology is also supported by findings that students usually want effective, supportive, and non-judgmental answers from chatbots, which makes AI technology serve as a source of comfort in a testing academic environment \citep{brandtzaeg2022, zhang2025}. The differences in how students use AI chatbots highlight the importance of understanding the factors causing these differences. Studies suggest that factors like stress, low confidence, and pressure can make some students rely on AI more often or in a more unhealthy way \citep{saglam2025, zhang2025}. These findings indicate that individual psychological characteristics may shape whether students adopt functional, emotional, or mixed ways of AI usage, highlighting the need for a theoretical framework to explain these differences.

\subsection{Human-AI vs Human-Human Interaction Patterns}

In spite of the fast development in the field of studies on human-AI chatbot communication, there are few theoretical studies that consider individual differences in human-AI chatbot communication, and the appropriateness of models developed in human-human communication to human-AI communication has been in doubt \citep{fox2021, skjuve2021}. Most studies target the cognitive aspects such as transparency and performance, and few explore the psychological processes involved in human-AI chatbot communication.

There is new evidence to show that human-AI interactions share many interpersonal traits with human-human interactions. These traits include vulnerability disclosure during emotionally relevant interactions \citep{skjuve2021}, intimacy simulation through human-like conversational characteristics \citep{adamopoulou2020}, the ability to demonstrate empathy and serve as a source of emotional support \citep{jiang2022}, and the tendency to have interactions with increasing proximity over time \citep{skjuve2021}. Based on this evidence, existing research suggests that human-AI interaction is indeed interpersonal and that models of human-human interaction can be employed to interpret human-AI interaction as well. Because attachment theory has long been used as an explanation of the manner in which people participate in relationships \citep{shaver1987}, and is essentially tied to the formation of trust and intimacy \citep{gillath2016, collins1990}, this theory is helpful to apply to the differences that exist between students interacting with AI-powered chatbots.

\subsection{Attachment Theory and AI Use}

Attachment theory, conceived by Bowlby to identify the evolution of emotional attachment in the relationship between infants and caregivers \citep{bowlby1969}, has been further developed in the adult population to understand patterns of romantic relationship attachment, loneliness, and loss coping strategies \citep{shaver1987}. In adult samples, attachment is commonly measured through the two factors of anxiety and avoidance \citep{brennan1998}. People with high attachment anxiety are very anxious about rejection and seek intimacy with extreme passion, and people with high avoidance prefer to maintain a distance and feel uncomfortable with intimacy \citep{shaver1987}. Persons low on both factors are known as securely attached and have a tendency to build trustworthy relationships \citep{shaver1987}.

Recent studies demonstrate not only the applicability of attachment theory in human-human relationships but also its generalization to humans' attachment to non-human entities, such as pet animals \citep{ciacchella2024, zilcha2011}. Since humans are able to treat non-human entities differently in light of their attachment to those entities, it is likely that humans might treat artificial intelligence in a similar manner. Previous research points to a definition of a three-component attachment construct that includes proximity-seeking behavior, a safe haven, or a secure base \citep{heffernan2012}. Studies also provide evidence that chatbots might serve to some degree in these capacities to produce emotional assistance, comfort in stressful conditions, and a feeling of being accompanied or heard \citep{adamopoulou2020, jiang2022, skjuve2021}. Together, these studies suggest that attachment theory might help to interpret the variations in relationships between college students and AI chatbots.

\subsection{Current Study}

Existing studies have yet to delve into the effects of attachment variation among college students on their interaction with AI, and the current literature provides no conclusive findings on how differently attached college students evaluate their interactions with a chatbot. The current study thus seeks to answer the following research question: \textit{How do college students with different attachment styles describe their interactions with an AI chatbot?}

\section{Method}

\subsection{Participants}

Seven undergraduate college students participated in this study ($N = 7$), comprising 5 females and 2 males, aged 21--22 years. Sampling methods included both convenience and snowball sampling. The selection criteria included being an active college student aged 18 or older, proficient in English, in a romantic relationship, and an active user of ChatGPT, having accessed the platform at least 2--3 times a week for the past three months. Purposive sampling was conducted to ensure diversity regarding attachment style, regardless of their field of study (Computer Science, Chemistry, Economics, Media \& Communications, and Psychology), as well as their year of study (junior \& senior). Participants were asked to self-report their attachment style after being briefed about attachment theory during the interview.

\subsection{Procedure}

Data collection took place in November 2025 using offline semi-structured interviews administered by 3 members of the team, who each interviewed 2 to 3 participants. All interviews were audio-recorded with participant consent and lasted anywhere from 24 to 38 minutes per interview. The interview guide consisted of two parts. The first part consisted of 7 questions developed from Experiences in Close Relationships-Short Form (ECR-S) \citep{wei2007}, rephrased as open-ended questions permitting qualitative responses about behaviors associated with attachment. The interviewer subsequently briefly described a four-category model classification of attachment theory \citep{bartholomew1991}: secure, anxious, avoidant, and fearful-avoidant styles. Participants then selected their own attachment style. The second part comprises five questions about observed behaviors and experiences associated with AI chatbots (see \hyperref[appendix:protocol]{Appendix A} for the interview protocol).

\subsection{Thematic Qualitative Coding}

Interview transcripts were analyzed using an inductive thematic analysis approach, involving 3 stages: open coding, axial coding, and selective coding. In open coding, individual coding by all team members entailed coding line by line and assigning labels to important text pieces. Axial coding involved comparing all participants' codes and grouping similar codes into categories according to conceptual similarity. In selective coding, categories were combined around the core code. To enhance inter-rater reliability, codes were assigned independently by coders and later compared and agreed upon after meetings to settle divergent codes, resulting in 18 codes and 4 categories, with 3 core themes (see \hyperref[appendix:codebook]{Appendix B} for the full codebook).

\subsection{Ethics Statement}

This study was conducted in accordance with ethical guidelines for research involving human participants. All participants provided informed consent prior to the interviews, and were informed of their right to withdraw from the study at any time without penalty. To protect participant privacy, all interview data were anonymized during transcription, and identifying information was removed from all reported quotations.

\section{Results}

Based on the self-reported attachment styles, 3 participants had a secure attachment, 3 had the avoidant style, and 1 had the mixed secure-avoidant style. Securely attached participants felt comfortable with emotional intimacy and preferred looking for support from their partners. The participants with avoidant attachment felt uneasy with vulnerability, preferred regulating their emotions on their own, and felt like they were burdens to their partners. Three themes emerged from the grounded theory analysis.

\subsection{Theme 1: AI as a Low-Risk Emotional Space}

Participants across attachment styles found the AI extremely safe to express their feelings, attributing it to the fact that the AI was non-judgmental, immediate, and light on interpersonal burden. An avoidant participant stated, ``There's no worry about saying the wrong thing or hurting the feelings of the AI,'' indicating the process was light on interpersonal burden. Another participant expanded on the benefit that the AI was ``instantly responsive to me and never makes me feel disappointed.''

\subsection{Theme 2: Attachment-Congruent Patterns of AI Engagement}

The patterns of AI use are consistent with individuals' attachment orientations. Participants who were securely attached used AI as an additional tool within the context of a larger support system, illustrated by the comment: ``I wouldn't replace conversations with her with conversations with an AI. It is more like a tool I use to help me get my thoughts together.'' Those classified as avoidant used the AI to buffer exposure to vulnerability, indicating that they ``generally wouldn't turn to my partner when my emotions are the highest, but the AI could always do that any time.'' One participant conceded that the AI might be used as a crutch: ``I might use the AI as a crutch so I don't have to be vulnerable to another human being.''

\subsection{Theme 3: The Paradox of AI Intimacy}

Some participants talked about disclosures made to AI but not partners, for example, ``I would tell ChatGPT things I would never tell him.'' This was seen as a consequence of lower stakes in AI relationships, as human relationships involve possible rejection, assessment, or derision. In turn, each participant had an awareness of limited AI capability, qualified by the recognition ``at the end of the day, it's still just a machine.'' This ambivalence reflects the notion that AI is used as a shield against vulnerability, rather than a direct tool for relationships.

\subsection{Summary}

These findings suggest that interactions college students have with AI chatbots are mediated by their attachment orientations. Students secure in their attachments tend to assimilate AI chatbots as an additional resource for an already present support system, while those who are avoidant may use the unique affordances provided by AI chatbots for emotion regulation with the goal of maintaining interpersonal boundaries. The core theoretical contribution---\textit{attachment-congruent AI chatbot use}---holds the critical insight that AI chatbot use is not a replacement for relational behavior. Instead, the chatbots provide an opportunity for the expression of pre-existing attachments in a new environment.

\section{Discussion}

The present study explored how college students with different attachment styles describe their interactions with ChatGPT, providing insights on the relational dimensions of human-AI interaction. Three main themes emerged: AI as a low-risk emotional space, attachment-congruent patterns of AI engagement, and the paradox of AI intimacy. These findings align with previous research showing that human-AI interactions resemble natural language processing seen in human-human interactions \citep{yigci2024}. More importantly, the results provided insight into how attachment style orientations shape the ways students experience and interpret their interactions with AI in the emotional domain.

Across attachment styles, participants described their conversations with AI as a safe and low-risk emotional space, consistent with prior research highlighting the non-judgmental characteristic of AI chatbots \citep{brandtzaeg2022, skjuve2021}. Students underscored that low interpersonal burden and potential emotional consequences contributed to making them feel safe when disclosing personal emotions and thoughts. This finding suggests that AI chatbots may be used as a safe space during moments of emotional uncertainty or psychological distress.

Moreover, students' patterns of interactions with AI closely reflect their human-human interactions, as evident in the theme of attachment-congruent AI-use. Securely attached participants described using AI as part of their support system as a reflective tool to better understand their thoughts and emotions, rather than a substitute for human connection. On the contrary, avoidantly attached participants primarily use AI as a tool to minimize their vulnerability within close relationships as they tend to manage emotional challenges independently. This aligns with prior research showing securely attached individuals are more open to relational support, whereas avoidantly attached prioritize self-reliance and emotional distance \citep{brennan1998, shaver1987}.

The final theme, the paradox of AI intimacy, captured the ambivalent nature of human-AI relationships. Interestingly, while recognizing that AI lacks genuine understanding of complex human experiences, participants described sharing conflicting thoughts with AI that they would not otherwise feel safe disclosing to their romantic partners. This tension suggests that AI may facilitate personal disclosure through providing a non-judgmental, low-risk environment, but it is not sufficient to fully replace the emotional aspects of genuine human connections. This resonates with concerns raised about the risks of over-reliance on AI as a coping mechanism \citep{saglam2025}. Together, these findings suggest that attachment orientation plays an important role in shaping students' interaction patterns with AI chatbots in ways that parallel their relational patterns in human relationships, highlighting meaningful similarities between human--human and human--AI interactions.

\subsection{Practical Implications}

These findings offer several practical implications for educators, mental health professionals, and AI developers. For \textit{educators and academic advisors}, understanding that students use AI chatbots not only for academic support but also for emotional regulation suggests a need for digital literacy curricula that address healthy AI use patterns. Educators should be aware that students with different attachment orientations may develop different relationships with AI tools, and some students---particularly those with avoidant attachment---may use AI to avoid seeking human support when it would be more beneficial.

For \textit{mental health professionals}, these findings highlight the importance of assessing clients' AI use patterns as part of understanding their broader relational functioning. Clinicians working with college students should consider how AI chatbot use may reflect or reinforce existing attachment patterns. For avoidantly attached individuals, AI may serve as a way to avoid vulnerability in human relationships, which could be addressed therapeutically. At the same time, AI chatbots could potentially be leveraged as a low-stakes space for practicing emotional expression before transferring these skills to human relationships.

For \textit{AI developers}, these results suggest that one-size-fits-all chatbot designs may not serve all users equally well. Developers might consider incorporating features that encourage users to seek human connection when appropriate, rather than positioning AI as a complete substitute for human support. Additionally, understanding that users may develop attachment-like patterns with AI systems carries ethical implications for how chatbots are designed and marketed, particularly to vulnerable populations.

\subsection{Limitations and Future Directions}

Several limitations of the present study should be acknowledged. First, the sample size was small ($N = 7$), which limits the generalizability of the findings to the broader population of college students. While small samples are common in qualitative research and allow for in-depth exploration, future studies should include larger and more diverse samples to validate these findings. Second, participants were recruited from a single university, which may limit the applicability of results to students from different institutional or cultural contexts. Third, attachment styles were assessed through self-report following a brief explanation, rather than using standardized measures such as the full ECR scale. This approach may have introduced measurement imprecision. Fourth, the study focused exclusively on ChatGPT; findings may not generalize to other AI chatbots with different interaction designs or personas.

Future research should address these limitations by employing larger, more diverse samples and validated attachment measures. Longitudinal studies could examine how attachment-AI interaction patterns evolve over time, particularly as users develop longer-term relationships with AI chatbots. Additionally, quantitative studies could test the themes identified here using survey methods to establish statistical relationships between attachment dimensions and specific AI use behaviors. Comparative studies across different AI platforms and cultural contexts would further enrich our understanding of attachment processes in human-AI interaction.

\section{Conclusion}

This study provides initial evidence that attachment theory offers a valuable framework for understanding individual differences in how college students interact with AI chatbots. The identification of attachment-congruent AI chatbot use as a core pattern suggests that students' pre-existing relational orientations shape their engagement with AI technologies in predictable ways. As AI chatbots become increasingly integrated into educational and personal contexts, understanding the psychological dimensions of these interactions becomes essential. These findings have implications for the design of AI systems, the development of digital literacy curricula, and clinical considerations regarding healthy AI use among young adults.

\section*{Author Contributions}

Both authors contributed equally to this work. \textbf{Ziqi Lin}: Conceptualization, Methodology, Investigation, Data Curation, Formal Analysis, Writing -- Original Draft, Writing -- Review \& Editing. \textbf{Taiyu Hou}: Conceptualization, Methodology, Investigation, Data Curation, Formal Analysis, Writing -- Original Draft, Writing -- Review \& Editing.

\section*{Declaration of Competing Interests}

The authors declare that they have no known competing financial interests or personal relationships that could have appeared to influence the work reported in this paper.

\section*{Data Availability}

The interview data generated during this study are not publicly available due to participant privacy and confidentiality concerns. The interview protocol (\hyperref[appendix:protocol]{Appendix A}) and codebook (\hyperref[appendix:codebook]{Appendix B}) used for analysis are provided in the appendices.

\bibliographystyle{apalike}
\bibliography{references}

@article{adamopoulou2020,
  author = {Adamopoulou, Eleni and Moussiades, Lefteris},
  title = {Chatbots: History, technology, and applications},
  journal = {Machine Learning with Applications},
  volume = {2},
  pages = {100006},
  year = {2020},
  doi = {10.1016/j.mlwa.2020.100006}
}

@book{bowlby1969,
  author = {Bowlby, John},
  title = {Attachment and loss: Vol. 1. Attachment},
  publisher = {Basic Books},
  year = {1969}
}

@article{bao2022,
  author = {Bao, Xiaofei and Li, Shuang and Zhang, Yi and Tang, Qun and Chen, Xu},
  title = {Different effects of anxiety and avoidance dimensions of attachment on interpersonal trust: A multilevel meta-analysis},
  journal = {Journal of Social and Personal Relationships},
  volume = {39},
  number = {7},
  pages = {2069--2093},
  year = {2022},
  doi = {10.1177/02654075221074387}
}

@article{bartholomew1991,
  author = {Bartholomew, Kim and Horowitz, Leonard M.},
  title = {Attachment styles among young adults: A test of a four-category model},
  journal = {Journal of Personality and Social Psychology},
  volume = {61},
  number = {2},
  pages = {226--244},
  year = {1991},
  doi = {10.1037/0022-3514.61.2.226}
}

@article{brandtzaeg2022,
  author = {Brandtzaeg, Petter Bae and Skjuve, Marita and Følstad, Asbjørn},
  title = {My {AI} friend: How users of a social chatbot understand their human--{AI} friendship},
  journal = {Human Communication Research},
  volume = {48},
  number = {3},
  pages = {404--429},
  year = {2022},
  doi = {10.1093/hcr/hqac008}
}

@incollection{brennan1998,
  author = {Brennan, Kelly A. and Clark, Catherine L. and Shaver, Phillip R.},
  title = {Self-report measurement of adult attachment: An integrative overview},
  booktitle = {Attachment theory and close relationships},
  editor = {Simpson, Jeffry A. and Rholes, W. Steven},
  pages = {46--76},
  publisher = {The Guilford Press},
  year = {1998}
}

@article{ciacchella2024,
  author = {Ciacchella, Chiara and Veneziani, Giulia and Garenna, Silvia A. and Lai, Carlo},
  title = {Interpersonal and pet bonding: A meta-analytic review of attachment dimensions},
  journal = {Journal of Social and Personal Relationships},
  volume = {42},
  number = {1},
  pages = {337--364},
  year = {2024},
  doi = {10.1177/02654075241285440}
}

@article{collins1990,
  author = {Collins, Nancy L. and Read, Stephen J.},
  title = {Adult attachment, working models, and relationship quality in dating couples},
  journal = {Journal of Personality and Social Psychology},
  volume = {58},
  pages = {644--663},
  year = {1990},
  doi = {10.1037/0022-3514.58.4.644}
}

@article{fageeh2025,
  author = {Fageeh, Abdulaziz},
  title = {The rise of chatbots in higher education: Exploring user profiles, motivations, and integration strategies},
  journal = {Social Sciences \& Humanities Open},
  volume = {12},
  pages = {101996},
  year = {2025},
  doi = {10.1016/j.ssaho.2025.101996}
}

@article{fox2021,
  author = {Fox, Jesse and Gambino, Andrew},
  title = {Relationship development with humanoid social robots: Applying interpersonal theories to human--robot interaction},
  journal = {Cyberpsychology, Behavior, and Social Networking},
  volume = {24},
  number = {5},
  pages = {294--299},
  year = {2021},
  doi = {10.1089/cyber.2020.0181}
}

@techreport{freeman2025,
  author = {Freeman, Josh},
  title = {Student generative {AI} survey 2025},
  institution = {Higher Education Policy Institute},
  type = {Policy Note},
  number = {61},
  year = {2025},
  url = {https://www.hepi.ac.uk/2025/02/26/student-generative-ai-survey-2025/}
}

@article{gallistl2024,
  author = {Gallistl, Martina and Kungl, Melanie and Gabler, Stefanie and Kanske, Philipp and Vrticka, Pascal and Engert, Veronika},
  title = {Attachment and inter-individual differences in empathy, compassion, and theory of mind abilities},
  journal = {Attachment \& Human Development},
  volume = {26},
  number = {4},
  pages = {350--365},
  year = {2024},
  doi = {10.1080/14616734.2024.2376762}
}

@book{gillath2016,
  author = {Gillath, Omri and Karantzas, Gery C. and Fraley, R. Chris},
  title = {Adult attachment: A concise introduction to theory and research},
  publisher = {Elsevier Inc.},
  year = {2016},
  doi = {10.1016/C2013-0-09705-8}
}

@article{heffernan2012,
  author = {Heffernan, Mary E. and Fraley, R. Chris and Vicary, Amanda M. and Brumbaugh, Claudia C.},
  title = {Attachment features and functions in adult romantic relationships},
  journal = {Journal of Social and Personal Relationships},
  volume = {29},
  number = {5},
  pages = {671--693},
  year = {2012},
  doi = {10.1177/0265407512443435}
}

@article{jiang2022,
  author = {Jiang, Qiaolei and Zhang, Yuxiang and Pian, Wenjing},
  title = {Chatbot as an emergency exit: Mediated empathy for resilience via human--{AI} interaction during the {COVID-19} pandemic},
  journal = {Information Processing \& Management},
  volume = {59},
  number = {6},
  pages = {103074},
  year = {2022},
  doi = {10.1016/j.ipm.2022.103074}
}

@article{klimova2025,
  author = {Klimova, Blanka and Pikhart, Marcel},
  title = {Exploring the effects of artificial intelligence on student and academic well-being in higher education: A mini-review},
  journal = {Frontiers in Psychology},
  volume = {16},
  pages = {1498132},
  year = {2025},
  doi = {10.3389/fpsyg.2025.1498132}
}

@article{labadze2023,
  author = {Labadze, Levan and Grigolia, Maya and Machaidze, Lela},
  title = {Role of {AI} chatbots in education: Systematic literature review},
  journal = {International Journal of Educational Technology in Higher Education},
  volume = {20},
  pages = {56},
  year = {2023},
  doi = {10.1186/s41239-023-00426-1}
}

@article{lin2025,
  author = {Lin, Haoran and Chen, Qiang},
  title = {Does artificial intelligence-assisted learning positively affect college students' motivation, emotion regulation, and academic uncertainty? Insights from situated learning theory},
  journal = {Learning and Motivation},
  volume = {92},
  pages = {102202},
  year = {2025},
  doi = {10.1016/j.lmot.2025.102202}
}

@article{macdonald2023,
  author = {Macdonald, Claudia and Adeloye, Davies and Sheikh, Aziz and Rudan, Igor},
  title = {Can {ChatGPT} draft a research article? An example of population-level vaccine effectiveness analysis},
  journal = {Journal of Global Health},
  volume = {13},
  pages = {01003},
  year = {2023},
  doi = {10.7189/jogh.13.01003}
}

@article{morell2025,
  author = {Morell-Mengual, Vicente and Fernández-García, Olga and Berenguer, Carmen and Ortega-Barón, Jessica and Gil-Llario, María Dolores and Estruch-García, Víctor},
  title = {Characteristics, motivations and attitudes of students using {ChatGPT} and other language model-based chatbots in higher education},
  journal = {Education and Information Technologies},
  volume = {30},
  pages = {22257--22274},
  year = {2025},
  doi = {10.1007/s10639-025-13650-1}
}

@article{saglam2025,
  author = {Sağlam, Reyhan Kıyak and Kalanlar, Burcu},
  title = {Living with and without {AI}: A mixed-methods study on {AI} usage, addiction, and '{AIlessphobia}' in nursing students},
  journal = {Nurse Education in Practice},
  volume = {88},
  pages = {104530},
  year = {2025},
  doi = {10.1016/j.nepr.2025.104530}
}

@article{shaver1987,
  author = {Shaver, Phillip and Hazan, Cindy},
  title = {Being lonely, falling in love: Perspectives from attachment theory},
  journal = {Journal of Social Behavior \& Personality},
  volume = {2},
  number = {2, Pt. 2},
  pages = {105--124},
  year = {1987}
}

@article{skjuve2021,
  author = {Skjuve, Marita and Følstad, Asbjørn and Fostervold, Knut Inge and Brandtzaeg, Petter Bae},
  title = {My chatbot companion - A study of human-chatbot relationships},
  journal = {International Journal of Human-Computer Studies},
  volume = {149},
  pages = {102601},
  year = {2021},
  doi = {10.1016/j.ijhcs.2021.102601}
}

@article{wei2007,
  author = {Wei, Meifen and Russell, Daniel W. and Mallinckrodt, Brent and Vogel, David L.},
  title = {The {Experiences in Close Relationship Scale (ECR)-Short Form}: Reliability, validity, and factor structure},
  journal = {Journal of Personality Assessment},
  volume = {88},
  number = {2},
  pages = {187--204},
  year = {2007},
  doi = {10.1080/00223890701268041}
}

@inproceedings{xie2022,
  author = {Xie, Tianling and Pentina, Iryna},
  title = {Attachment theory as a framework to understand relationships with social chatbots: A case study of {Replika}},
  booktitle = {Proceedings of the 55th Hawaii International Conference on System Sciences},
  year = {2022},
  doi = {10.24251/hicss.2022.258}
}

@article{yigci2024,
  author = {Yigci, Derin and Eryilmaz, Melih and Yetisen, Ali K. and Tasoglu, Savas and Ozcan, Aydogan},
  title = {Large language model-based chatbots in higher education},
  journal = {Advanced Intelligent Systems},
  volume = {7},
  number = {3},
  pages = {2400429},
  year = {2024},
  doi = {10.1002/aisy.202400429}
}

@misc{zhang2025,
  author = {Zhang, Yue and Zhao, Dong and Hancock, Jeffrey T. and Kraut, Robert and Yang, Diyi},
  title = {The rise of {AI} companions: How human--chatbot relationships influence well-being},
  year = {2025},
  eprint = {2506.12605},
  archiveprefix = {arXiv},
  primaryclass = {cs.HC},
  doi = {10.48550/arXiv.2506.12605}
}

@article{zilcha2011,
  author = {Zilcha-Mano, Sigal and Mikulincer, Mario and Shaver, Phillip R.},
  title = {An attachment perspective on human--pet relationships: Conceptualization and assessment of pet attachment orientations},
  journal = {Journal of Research in Personality},
  volume = {45},
  number = {4},
  pages = {345--357},
  year = {2011},
  doi = {10.1016/j.jrp.2011.04.001}
}

\vspace{2em}
\appendix

\section{Semi-Structured Interview Protocol}
\label{appendix:protocol}

\subsection*{Part 1: Human Relationships and Attachment}
\textit{Adapted from ECR-S \citep{wei2007}}

\begin{enumerate}
    \item When you feel stressed or need emotional support, do you usually turn to your romantic partner? Why or why not?
    \item How do you typically feel when your partner is not immediately available to you?
    \item Do you ever worry that your partner may not care about the relationship as much as you do?
    \item How comfortable are you with emotional closeness in a romantic relationship?
    \item When a relationship becomes very close, do you ever feel the urge to pull back or create distance?
    \item How do you usually handle sharing your feelings or personal concerns with a partner?
    \item Based on what you just learned about different attachment styles, which style most accurately describes you?
\end{enumerate}

\subsection*{Part 2: AI Chatbot Use}

\begin{enumerate}
    \item How often do you use AI Chatbots? How do you usually use AI chatbots?
    \item Have you ever used AI chatbots during stress, confusion, or emotional moments?
    \item If you had to describe your relationship with AI chatbots, how would you describe it?
    \item In what ways does interacting with an AI chatbot feel similar to or different from interacting with a person?
    \item Do you feel your interactions with AI chatbots stay on the surface or sometimes become deeper?
\end{enumerate}

\section{Codebook}
\label{appendix:codebook}

\footnotesize
\begin{longtable}{p{2cm}p{1.8cm}p{2.4cm}p{5.8cm}}
\toprule
\textbf{Theme} & \textbf{Category} & \textbf{Code} & \textbf{Definition} \\
\midrule
\endfirsthead
\toprule
\textbf{Theme} & \textbf{Category} & \textbf{Code} & \textbf{Definition} \\
\midrule
\endhead
\midrule
\multicolumn{4}{r}{\textit{Continued on next page}} \\
\endfoot
\bottomrule
\endlastfoot

Attachment Patterns & Self-Protection & Fear of burdening & Concern about overwhelming a partner with emotional needs \\
 & & Avoidance of vulnerability & Reluctance to show insecure or messy parts to a partner \\
 & & Maintaining image & Desire to appear capable and put-together to partner \\
\cmidrule{2-4}
 & Relational Comfort & Comfort with closeness & Ease with emotional intimacy in relationships \\
 & & Distance-seeking & Urge to pull back when the relationship intensifies \\
\midrule
AI Engagement & Functional Use & AI as an academic tool & Using AI for coursework, coding, and information retrieval \\
\cmidrule{2-4}
 & Emotional Use & AI for emotional support & Using AI to process stress, anxiety, or emotions \\
 & & AI for pre-processing & Using AI to organize thoughts before a human conversation \\
\midrule
AI Affordances & Safety Features & Non-judgmental space & AI does not judge or remember negatively \\
 & & Low interpersonal burden & No worry about burdening AI or saying the wrong things \\
 & & No risk/consequences & AI interaction has no relational stakes \\
\cmidrule{2-4}
 & Availability & AI immediacy & AI is always available and responds instantly \\
\midrule
Human-AI Comparison & Differential Disclosure & Easier vulnerability with AI & More honest/open with AI than with partner \\*
\cmidrule{2-4}
 & Relational Role & AI as a supplement & AI complements but does not replace human relationships \\*
 & & AI as a substitute & AI is used instead of unavailable humans \\*
\cmidrule{2-4}
 & AI Limitations & AI lacks genuine connection & Recognition that AI cannot truly understand or care \\
\bottomrule
\end{longtable}
\normalsize

\end{document}